\documentclass[12pt]{article}
\usepackage[a4paper,top=1.5in,textwidth=17cm,textheight=23cm,twoside]{geometry}

\usepackage{fancyhdr}
\usepackage{copyright_instance}

\usepackage{hyperref} %%lipics%% ce l'ha gia'
\usepackage{breakurl}
\usepackage{amssymb,amsmath}
\usepackage{mathrsfs}    %%% per \mathscr{A} di automa
\usepackage{stmaryrd}    %%% per \llbracket \rrbracket
\usepackage[ruled,vlined]{algorithm2e}%boxed,longend
\usepackage{color}
\usepackage[all,dvips,arc,curve,color,frame]{xy}  %%% buttare in seguito a generazione ps, pdf
\CompileMatrices                                  %%% buttare in seguito a generazione ps, pdf
\usepackage{graphicx}
\usepackage{amssymb,amsmath,amsthm}
\SetKwInOut{Input}{input}
\SetKwInOut{Output}{output}
\SetKwInOut{DataStructures}{data structures}
\AlgoDontDisplayBlockMarkers\SetAlgoNoEnd\SetAlgoNoLine%%%%%% da ripetere localmente
\newtheorem{theorem}{Theorem}[section]

\newtheorem{definition}[theorem]{Definition}
\SetKwProg{Fun}{function}{}{end}

\definecolor{codeCol}{rgb}{0.4,0,0.4}
\newcommand{\bcode}{\color{codeCol}}
\newcommand{\pqProofForget}[1]{}%#1}
\newcommand{\pqTextForget}[1]{}
\newcommand{\pqLongText}[1]{}

\newcommand{\forget}[1]{}

\newcommand{\acapo}{\\}

\newcommand{\dotInItem}{\cdot}
\newcommand{\set}[1]{\mbox {$\{#1\}$}}
\newcommand{\indexedset}[2]{{\{#1\}_{#2}}}
\newcommand{\pair}[2]{(#1 , #2 )}
\newcommand{\lookEquation}[2]{\mbox {$#1 \stackrel{.}{=} #2$}}
\newcommand{\sItem}[2]{\mbox {$[ #1 , #2 ]$}}

\newcommand{\size}[1]{\mathop{\mid} #1 \mathop{\mid}}

\newcommand{\Nzeroitem}{LR(0)-item}
\newcommand{\Noneitem}{LR(1)-item}
\newcommand{\Nzeroitems}{LR(0)-items}
\newcommand{\Noneitems}{LR(1)-items}
\newcommand{\NvarSet}{\mbox {$\,\mathbb{V}$}} %%% \mathcal{V}ar$}}
\newcommand{\varX}{x}
\newcommand{\varXp}{x'}

\newcommand{\Cvars}{V\!ar\!s}
\newcommand{\RCvars}{RV\!ar\!s}
\newcommand{\CvarsReducing}{V\!ar\!s_r}
\newcommand{\CvarsBypassing}{V\!ar\!s_b}
\newcommand{\Cequations}{Eqs}
\newcommand{\CequationsBypassing}{Eqs_b}
\newcommand{\RCequations}{R\!Eqs}
\newcommand{\Cclass}[1]{class(#1)}

\newcommand{\assign}{\longleftarrow}

\newcommand{\FnewVar}{\mbox {${\rm{newVar}}()$}}
\newcommand{\Fclosure}[1]{\mbox {${\rm{closure}} (#1)$}}
\newcommand{\Ffirst}[1]{\mbox {${\rm{first}} (#1)$}}
\newcommand{\Ffollow}[1]{\mbox {${\rm{follow}} (#1)$}}
\newcommand{\Fkernel}[1]{\mbox {${\rm{kernel}} (#1)$}}
\newcommand{\Fgoto}{\mbox {$\tau_{s}$}}%%{\mbox {${\rm{goto}}$}}

\newcommand{\Fdequeue}[1]{\mbox {${\rm{dequeue}} (#1)$}}
\newcommand{\Fproj}[1]{\mbox {${\rm{prj}} (#1)$}}
\newcommand{\Token}{\mbox {${\rm{id}}$}}
\newcommand{\Fla}{\mathcal{L}\mathcal{A}}
\newcommand{\Fground}[1]{\mbox {${\rm{ground}}(#1)$}}

\newcommand{\Ftraverse}[1]{\mbox {${\rm{search}} (#1)$}}

\newcommand{\yacc}{{\texttt{Yacc}}}
\newcommand{\Ftrans}[1]{\mathrel{\stackrel{#1}{\longmapsto}}}

\newcommand{\LRm}{LRm(1)}

\newcommand{\statesX}{St_{s}}

\newcommand{\autL}{{\mathscr{A}}_l}

\newcommand{\autM}{{\mathscr{A}}_m}

\newcommand{\Fval}[1]{val(#1)}

\newcommand{\FclosureOne}[1]{\mbox {${\rm{closure}}_1 (#1)$}}
\newcommand{\FclosureZero}[1]{\mbox {${\rm{closure}}_0 (#1)$}}
\newcommand{\pqcomment}[1]{}
\renewcommand{\Fgoto}{\mbox {$\tau$}}

\renewcommand{\statesX}{\mathcal{Q}}
\newcommand{\Ntmp}{\mbox {$T\!mp$}}
\renewcommand{\LRm}{LRm(1)}
\newcommand{\bison}{{\texttt{Bison}}}

\title{
Briefly on Bottom-up
}

\author{Paola Quaglia\\ {\small{University of Trento}}}
\date{\mbox{}}

\begin{document}

\maketitle

\subsection*{Abstract}
These short notes are meant as a quick reference for
the construction of SLR(1), of LR(1), and of LALR(1) parsing tables.

\

\noindent
{\fontsize{9.5}{12.5}\sffamily\bfseries {1998 ACM Subject Classification}}
{\fontsize{9.5}{12.5}\sffamily {F.4.2 Grammars and Other Rewriting Systems}}

\noindent
{\fontsize{9.5}{12.5}\sffamily\bfseries {Keywords and phrases}}
{\fontsize{9.5}{12.5}\sffamily {SLR(1) grammars; LR(1) grammars; LALR(1) grammars}}

\section{Outline}
We provide descriptions and references relative to the construction of
parsing tables for SLR(1), for LR(1), and for LALR(1) grammars.
The report is organized as follows.
Basic definitions and conventions are collected in Sec.~\ref{sec:notation}.
SLR(1), LR(1), and LALR(1) grammars are the subjects
of Sec.~\ref{sec:slr},
of Sec.~\ref{sec:lr}, and
of Sec.~\ref{sec:lalr}, respectively.
For grammars in each of these three classes, the construction of the
relative parsing tables is presented as an instance of a single schema.
The schema itself is described beforehand in Sec.~\ref{sec:caapt}.

\pqcomment{The algorithms referred to in these notes are all collected in the Appendix.
}% end\pqcomment

\section{Notation and basic definitions}
\label{sec:notation}

Basic definitions and notational conventions are summerized below.

A context-free grammar is a tuple $\mathcal{G} = (V,T,S,\mathcal{P})$,
where the elements of the tuple represent, respectively,
the vocabulary of terminal and nonterminal symbols,
the set of terminal symbols,
the start symbol,
and the set of productions.
Productions have the shape
$A\rightarrow\beta$ where $A\in{V\setminus{T}}$ is called the \emph{driver},
and $\beta\in{V^*}$ is called the \emph{body}.
The one-step rightmost derivation relation is denoted by
``$\Rightarrow$", and ``$\Rightarrow^*$" stands for
its reflexive and transitive closure.
A grammar is said \emph{reduced} if it does not contain any useless production,
namely any production that is never involved in the derivation of strings of
terminals from $S$.
We assume grammars be reduced.

\pqcomment{
  The size of a grammar is given by
  $\sum_{A\rightarrow\beta \in \mathcal{P} \size{A\alpha}}$,
  where $\size{\alpha}$ denotes the length of $\alpha$.
}%end\pqcomment

The following notational conventions are adopted.
The empty string is denoted by $\epsilon$.
Lowercase letters early in the Greek alphabet stand for strings of grammar symbols
  ($\alpha,\beta,\ldots \in V^*$),
lowercase letters early in the alphabet stand for terminals
  ($a,b,\ldots \in T$),
uppercase letters early in the alphabet stand for nonterminals
  ($A,B,\ldots \in (V \setminus T)$),
uppercase letters late in the alphabet stand for either terminals or nonterminals
  ($X,Y,\ldots \in V$), and
strings of terminals, i.e. elements of $T^*$, are ranged over by
      $w,w_0,\ldots$.

For every $\alpha$,
$\Ffirst{\alpha}$ denotes the set of terminals that begin strings $w$
such that $\alpha \Rightarrow^* w$.
Moreover, if $\alpha \Rightarrow^* \epsilon$ then $\epsilon\in\Ffirst{\alpha}$.
For every $A$,
$\Ffollow{A}$ denotes the set of terminals that can follow $A$ in a derivation,
and is defined in the usual way.

Given any context-free grammar $\mathcal{G}$,
parsing is applied to strings followed by the symbol $\$\notin V$ used as endmarker.
Also, the parsing table is produced for an enriched version of $\mathcal{G}$,
denoted by $\mathcal{G}' = (V',T,S',\mathcal{P}')$.
The enriched grammar $\mathcal{G}'$
is obtained from $\mathcal{G}$ by augmenting $V$ with a fresh nonterminal symbol $S'$,
and by adding the production $S' \rightarrow S$ to $\mathcal{P}$.

An \Nzeroitem\ of $\mathcal{G}'$ is a production of $\mathcal{G}'$ with the
distinguished marker ``$\dotInItem$"
at some position of its body,
like, e.g., $A\rightarrow\alpha\dotInItem\beta$.
The single \Nzeroitem\ for a production of the shape $A\rightarrow\epsilon$
takes the form $A\rightarrow\dotInItem$.
%%%
\pqcomment{
  An \Nzeroitem\ is called \emph{reducing} if it has the form
  $A\rightarrow\beta\dotInItem$.
  An LR(1)-item is called \emph{reducing} if its first component is a reducing
  \Nzeroitem.
  The \Nzeroitem\ ${A \rightarrow \alpha\dotInItem\beta}$ is called
  \emph{kernel item} if either $\alpha\neq\epsilon$ or $A = S'$,
  \emph{closure item} if it is not kernel,
  \emph{reducing item} if $\beta=\epsilon$, and
  \emph{bypassing item} if it is not reducing.
}%% end\pqcomment
The \Nzeroitems\
${S' \rightarrow \dotInItem{S}}$ and
${S' \rightarrow {S}\dotInItem}$ are called, respectively,
\emph{initial item} and \emph{accepting item}.
The \Nzeroitem\
${A \rightarrow \alpha\dotInItem\beta}$ is called
\begin{itemize}
\item
  \emph{kernel item} if it is either initial or such that $\alpha\neq\epsilon$,
\item
  \emph{closure item} if it is not kernel, and
\item
  \emph{reducing item} if it is not accepting and if $\beta=\epsilon$.
\end{itemize}

For a set of \Nzeroitems\ $P$,
$\Fkernel{P}$ is the set of the kernel items in $P$.
By definition, the initial item is the single kernel item of $\mathcal{G}'$
with the dot at the leftmost position, and
items of the shape $A\rightarrow\dotInItem$ are the only non-kernel
reducing items.

An \Noneitem\ of $\mathcal{G}'$ is a pair consisting of an
\Nzeroitem\ of $\mathcal{G}'$ and of a subset of $T\cup\set{\$}$,
like, e.g.,
$\sItem{A\rightarrow\alpha\dotInItem\beta}{\set{a,\$}}$.
The second component of an \Noneitem\ is
called lookahead-set and is ranged over by
$\Delta, \Gamma, \ldots$.
An \Noneitem\ is said \emph{initial}, \emph{accepting}, \emph{kernel},
\emph{closure} or \emph{reducing}
if so is its first component.
For a set $P$ of \Noneitems,
$\Fproj{P}$ is the set of \Nzeroitems\ occurring as first components of the
elements of $P$.
Also, function $\Fkernel{\_}$ is overloaded, so that for a set of
\Noneitems\ $P$,
$\Fkernel{P}$ is the set of the kernel items in $P$.

\section{Characteristic automata and parsing tables}
\label{sec:caapt}

Given a context-free grammar $\mathcal{G}$ and a string of terminals $w$, the
aim of bottom-up parsing is to deterministically reconstruct,
in reverse order and while
reading $w$ from the left, % and one symbol at a time,
a rightmost derivation of $w$ if the string belongs to the language generated by
$\mathcal{G}$.
If $w$ does not belong to the language, then parsing returns an error.
The computation is carried over on a stack, and before terminating with success or
failure, it consists in \emph{shift} steps and in \emph{reduce} steps.
A shift step amounts to pushing onto the stack the symbol of $w$ that is
currently pointed by the input cursor, and then advancing the cursor.
Each reduce step is relative to a specific production of $\mathcal{G}$.
A reduce step under $A\rightarrow\beta$ consists in popping $\beta$
off the stack and then pushing $A$ onto it.
Such reduction
is the appropriate kind of move when, for some $\alpha$ and $w_1$,
the global content of the stack is $\alpha\beta$ and
the rightmost derivation of the analyzed string $w$ takes the form
\begin{equation}
  \label{eq:forKnuthLong}
  S \Rightarrow^* \alpha A w_1 \Rightarrow \alpha \beta w_1 \Rightarrow^* w.
\end{equation}
A seminal result by Knuth~\cite{Knuth65} is that for reduced grammars
the language of the \emph{characteristic strings}, i.e. of the strings like
$\alpha\beta$ in (\ref{eq:forKnuthLong}),
is a regular language.
By that, a deterministic finite state automaton can be defined and used as the
basis of the finite control
of the parsing procedure~\cite{DeRemer69,DeRemer71}.
This automaton is referred to as the \emph{characteristic automaton}, and
is at the basis of the construction of the actual
controller of the parsing algorithm, the so-called
\emph{parsing table}.

If $\mathcal{Q}$ is the set of states of the characteristic automaton, then
the parsing table is a matrix $\mathcal{Q} \times (V \cup \set{\$})$,
and the decision about which step to take next depends on the current state
and on the symbol read from the parsed word.
Various parsing techniques use the same shift/reduce algorithm
but are driven by different controllers, which in turn are built on top of
distinct characteristic automata.

States of characteristic automata are sets of items.
A state $P$ contains the item $A\rightarrow\alpha\dotInItem\beta$
(or an item whose first projection is $A\rightarrow\alpha\dotInItem\beta$)
if $P$ is the state reached after recognizing a portion of the parsed word
whose suffix corresponds to an expansion of $\alpha$.
Each state of the characteristic automaton is generated from a kernel
set of items by closing it up to include all those items that,
w.r.t. the parsing procedure, represent the same progress as
that expressed by the items in the kernel.

The transition function $\tau$ of the automaton describes the evolution
between configurations.
Every state has as many transitions as the number of distinct symbols that
follow the marker ``$\dotInItem$" in its member items.
Assume the parser be in state $P_n$, and let $a$ be the current symbol read from
the parsed word.
If the entry $\pair{P_n}{a}$ of the parsing table is a shift move,
then the control goes to the state $\tau(P_n,a)$.
If it is a reduction move under $A\rightarrow\beta$,
then the next state is $\tau(P,A)$ where $P$ is the origin
of the path spelling $\beta$ and leading to $P_n$.
Precisely, suppose that $\beta=Y_1\ldots{Y_n}$ and let
$P \Ftrans{Y} P'$ denote that
$\tau(P,Y)=P'$.
Then the state of the parser after the reduction $A\rightarrow\beta$ in $P_n$
is $\tau(P,A)$ where
$P$ is such that
$P \Ftrans{Y_1} P_1 \Ftrans{Y_2} \ldots \Ftrans{Y_n} P_n$.

The common features of the various characteristic automata used
to construct bottom-up parsing tables are listed below.
\begin{itemize}
  \item
  Each state in $\mathcal{Q}$ is a set of items.
  \item
  The initial state contains the initial item.
  \item
  The set $\mathcal{F}$ of final states consists of all the states
  containing at least one reducing item.
  \item
  The vocubularly is the same as the vocabulary of the given grammar,
  so that the transition function takes the form
  $\tau:(\mathcal{Q} \times V) \rightarrow \mathcal{Q}$.
\end{itemize}

In the shift/reduce algorithm,
the decision about the next step depends on the
current configuration of the parser and on the current input terminal.
So, in order to set up a parsing table,
it is also necessary to define, for each final state $Q$
and for each reducing item in $Q$, which set of terminals
should trigger the relative reduction.
This is achieved by providing an actual
definition of the \emph{lookahed function}
%\begin{center}
  $\Fla: \mathcal{F} \times \mathcal{P} \rightarrow \wp(V\cup\set{\$}) $.
%\end{center}
For the argument pair $\pair{P}{A\rightarrow\beta}$ the lookahead function
returns the set of symbols calling for a reduction after
$A\rightarrow\beta$ when the parser is in state $P$.
E.g., referring to (\ref{eq:forKnuthLong}) and assuming that $P$ is the state
of the parser when $\alpha\beta$ is on the stack,
$\Fla\pair{P}{A\rightarrow\beta}$
is expected to contain the first symbol of $w_1$.

Once the underlying characteristic automaton and lookahead function are defined,
the corresponding parsing table is obtained as described below.

\begin{definition}
\label{def:parsing_table}
  Let $\mathcal{Q}$, $V$, and $\tau$ be, respectively, the set of states, the vocabulary,
  and the transition function of a characteristic automaton.
  Also, let $\Fla_i$ be an actual instance of the lookahead function.
  Then, the {\rm parsing table for} the pair constisting
  of the given characteristic automaton and
  of the given lookahead function
  is the matrix $\mathcal{Q} \times (V \cup \set{\$})$
  obtained by filling in each entry $(P,Y)$ after the following rules.
  \begin{itemize}
  \item
    Insert \emph{``Shift $Q$''}
    if $Y$ is a terminal and $\tau(P,Y)=Q$.
  \item
    Insert \emph{``Reduce $A \rightarrow \beta$''}
    if $P$ contains a reducing item for $A \rightarrow \beta$
    and $Y\in{\Fla}_i\pair{P}{A\rightarrow\beta}$.
  \item
    Set to \emph{``Accept''}
    if $P$ contains the accepting item and $Y=\$$.
  \item
    Set to \emph{``Error"}
    if $Y$ is a terminal or $\$$, and none of the above applies.
  \item
    Set to \emph{``Goto $Q$''}
    if $Y$ is a nonterminal and $\tau(P,Y)=Q$.
  \end{itemize}
\end{definition}

%\input{Inputs/alg_generic_automaton}

%%% alg_automaton
%% pag.32 del manuale: \eIf(then comment){condition}{then block}(else comment){else block}

%\IncMargin{1em}
\begin{algorithm*}%[H]

    initialize $\mathcal{Q}$ to contain $P_0$\;
    tag $P_0$ as unmarked\;

      \While{ there is an unmarked state $P$ in $\mathcal{Q}$
      }{
        mark $P$ \;
        \ForEach{ $Y$ on the right side of the marker in some item of $P$
        }{
          Compute in $\Ntmp$ the kernel-set of the $Y$-target of $P$\;

          \BlankLine
          \eIf{ $\mathcal{Q}$ already contains a state $Q$ whose kernel is $\Ntmp$
              }{
                \BlankLine
                Let $Q$ be the $Y$-target of $P$\;
              }{
                \BlankLine
                Add $\Fclosure{\Ntmp}$ as an unmarked state to the collection $\mathcal{Q}$\;
                \BlankLine
                Let $\Fclosure{\Ntmp}$ be the $Y$-target of $P$\;
              }
        }
      }
\caption{\label{alg:generic-automaton}
         Construction of LR(0)-automaton and of LR(1)-automaton\\
         ($P_0$, $\Ntmp$, and $\Fclosure{\_}$ to be instantiated accordingly)
        }
\end{algorithm*}
%\DecMargin{1em}

The table might have multiply-defined entries,
mentioning either a shift and a reduce directive (known as a shift/reduce conflict),
or multiple reduce directives for different productions (known as a reduce/reduce conflict).
If so, then the constructed
table cannot possibly drive a deterministic parsing procedure.
Consequently, grammar $\mathcal{G}$ is said not to belong to the class of grammars
syntactically analyzable by the methodology
(choice of automaton and of lookahead function)
underlying the definition of the parsing table.
Viceversa, if the constructed parsing table contains no conflict, then $\mathcal{G}$
belongs to the class of grammars parsable by the chosen methodology.

Below we focus on SLR(1) grammars, LR(1) grammars, and LALR(1) grammars.
Seen as classes of grammars,
SLR(1) is strictly contained in LALR(1) which is strictly contained in LR(1).

Some of the algorithms reported in the following are run on the grammar
$\mathcal{G}_1$ below, which is taken from~\cite{AhoU77}.
The language generated by $\mathcal{G}_1$ can be thought of as a language of assignments
of r-values to l-values, where an l-value can denote the content of an r-value.
Interestingly, $\mathcal{G}_1$ separates the class SLR(1) from the
class LALR(1).

\begin{center}
\label{page:g1}
$\begin{array}{llcl}
  \mathcal{G}_1:   & S & \rightarrow & L=R \mid R       \\
                   & L   & \rightarrow & *R \mid \Token   \\
                   & R   & \rightarrow & L
\end{array}$
\end{center}

\section{SLR(1) grammars}
\label{sec:slr}

\IncMargin{1em}
\begin{algorithm}[H]
  %\LinesNumbered
  %
  \BlankLine
    \Fun{$\FclosureZero{P}$
    }{

    tag every item in $P$ as unmarked \;

    \While{ there is an unmarked item $I$ in $P$
    }{
        mark $I$ \;

        \If{ $I$ has the form ${A \rightarrow\alpha\dotInItem{B}\beta}$
        }{

            \ForEach{${B\rightarrow\gamma} \in \mathcal{P}'$
            }{

              \If{ ${B\rightarrow\dotInItem\gamma} \notin {P}$
              }{
                  add ${B\rightarrow\dotInItem\gamma}$ as an unmarked item to $P$ \;
              }
            }
        }
    }

    \Return $P$ \;
    }
\caption{\label{alg:closureZero}
         Computation of $\FclosureZero{P}$
        }
\end{algorithm}
\DecMargin{1em}

The SLR(1) parsing table for $\mathcal{G}$ is constructed from an automaton,
called LR(0)-automaton,
whose states are sets of \Nzeroitems.
Correspondingly, function ${\Fla_i}$ is instantiated as follows.
\begin{quote}
\begin{em}
  For every final state $P$ of the LR(0)-automaton and
  for every $A\rightarrow\beta\dotInItem \in P$,
  \acapo
  $\Fla_{SLR}\pair{P}{A\rightarrow\beta} = \Ffollow{A}$.
\end{em}
\end{quote}

\noindent
LR(0)-automata are obtained by applying Alg.~\ref{alg:generic-automaton}
after:
\begin{itemize}
  \item
  using $\FclosureZero{\_}$ (see Alg.~\ref{alg:closureZero}) as
  $\Fclosure{\_}$ function, and
  \item
  taking $P_0 = \FclosureZero{\set{S'\rightarrow\dotInItem{S}}}$, and
  \item
  taking $\Ntmp =
          \set{{A\rightarrow\alpha{Y}\dotInItem\beta} \mid
               {A\rightarrow\alpha\dotInItem{Y}\beta}\in{P}
               \mbox{ for some } A,\alpha, \mbox{ and } \beta}$.
\end{itemize}

\noindent
The intuition behind the definition of $\FclosureZero{\_}$ is that,
if the parsing procedure progressed as encoded by
${A\rightarrow\alpha\dotInItem{B}\beta}$, and
if ${B\rightarrow\gamma} \in \mathcal{P}'$,
then the coming input can be an expansion of
$\gamma$ followed by an expansion of $\beta$.
In fact, $\FclosureZero{P}$ is defined as
the smallest set of items that satisfies the following equation:
\begin{quote}
  $\FclosureZero{P} =
    P \mathrel{\cup}
    \set{{B\rightarrow\dotInItem\gamma}
         \mbox{ such that }
        {A \rightarrow\alpha\dotInItem{B}\beta}\in\FclosureZero{P} \mbox{ and }
        {B\rightarrow\gamma}\in{\mathcal{P}'}
    }.
$
\end{quote}

As an example of application of Alg.~\ref{alg:closureZero}, the
items belonging to
$\FclosureZero{\set{S'\rightarrow\dotInItem{S}}}$
for $\mathcal{G}_1$ are shown below.

\begin{center}

$\begin{array}{l@{\hspace{2ex}}l}
  \FclosureZero{\set{S'\rightarrow\dotInItem{S}}}:
  &
  {S'\rightarrow\dotInItem S}
  \\ &
  {S\rightarrow\dotInItem L=R}
  \\ &
  {S\rightarrow\dotInItem R}
  \\ &
  {L\rightarrow\dotInItem *R}
  \\ &
  {L\rightarrow\dotInItem \Token}
  \\ &
  {R\rightarrow\dotInItem L}
\end{array}$
\end{center}

\noindent
The rationale for Alg.~\ref{alg:generic-automaton} is the following.
\begin{itemize}
  \item
    Compute the set of states of the automaton by starting from the initial state
    $P_0$ and incrementally adding the targets, under possible $Y$-transitions,
    of states already collected.
  \item
    To decide which, if any, is the $Y$-target of a certain state $P$,
    first compute in \Ntmp\ the set of the kernel items of the $Y$-target.
  \item
    Compare \Ntmp\ to the states in the current collection.
    If, for some collected $Q$, \Ntmp\ and $Q$ have the same
    kernel items, then take $Q$ as the $Y$-target of $P$.
    If no match is found for \Ntmp, then add $\FclosureZero{\Ntmp}$ to the current
    collection of states.
\end{itemize}

\begin{figure}[!h]
  \centering
  \includegraphics[width=14cm]{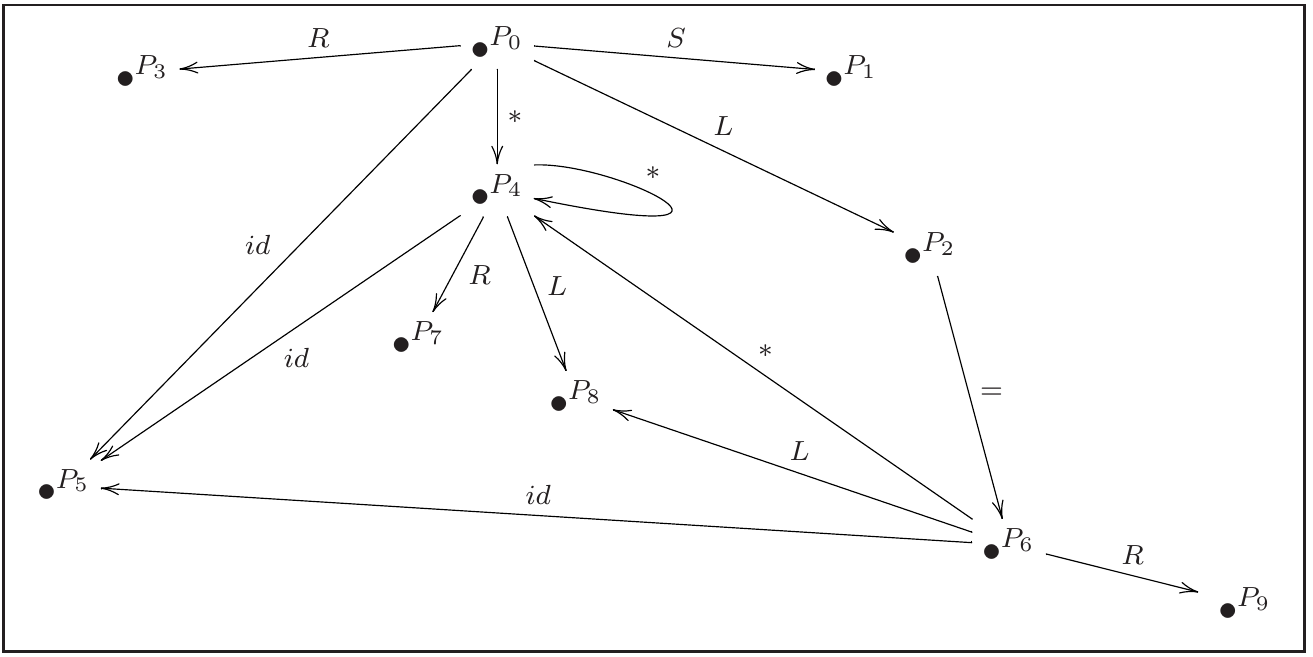}
  \caption{\label{fig:appendix_pointer_automaton}
         Layout of the LR(0)-automaton for $\mathcal{G}_1$
         }
\end{figure}

The layout of the LR(0)-automaton for $\mathcal{G}_1$ is reported in
Fig.~\ref{fig:appendix_pointer_automaton}.
The accepting item is in state $P_1$.
The final states of the automaton, and the reducing items they contain,
are listed below.

\begin{center}
$\begin{array}{l@{\hspace{4ex}}l}
  \mbox{{\bf State}}      &
  \mbox{{\bf Reducing item}}
\\[.7ex]
  P_2:       &
  {R\rightarrow L\dotInItem}
\\[.7ex]
  P_3:       &
  {S\rightarrow R\dotInItem}
\\[.7ex]
  P_5 :       &
  {L\rightarrow \Token \dotInItem}
\\[.7ex]
  P_7 :       &
  {L\rightarrow * R \dotInItem}
\\[.7ex]
  P_8 :       &
  {R\rightarrow L\dotInItem}
\\[.7ex]
  P_9 :       &
  {S\rightarrow L=R\dotInItem}
\end{array}$
\end{center}

$\mathcal{G}_1$ is not SLR(1).
Indeed, the SLR(1) parsing table for $\mathcal{G}_1$
has a shift/reduce conflict at the entry $\pair{P_2}{=}$.
This is due to the fact that $P_2$ has an outgoing transition labelled by $=$
(which induces a shift to $P_6$), and to the fact that
${=} \in \Ffollow{R}$
(which induces a reduce after ${R\rightarrow L}$).

\pqcomment{
  \noindent
  The upper bound to the number of states of the LR(0)-automaton for $\mathcal{G}$ is
  $O(2^{n})$, where $n$ is the size of $\mathcal{G}'$.
  This bound is tight.
  A grammar that shows it is reported in Ex.~4.6.7, pag.~258,
  of the international edition of~\cite{AhoLSU06}.
}% end\pqcomment

\section{LR(1) grammars}
\label{sec:lr}

\IncMargin{1em}
\begin{algorithm}%[H]
  %\LinesNumbered
  %
  \BlankLine
    \Fun{$\FclosureOne{P}$
    }{

      tag every item in $P$ as unmarked \;

      \While{ there is an unmarked item $I$ in $P$
      }{
          mark $I$ \;

          \If{ $I$ has the form $\sItem{A \rightarrow\alpha\dotInItem{B}\beta}{\Delta}$
          }{
              $\Delta_1 \longleftarrow \bigcup_{d\in \Delta}{\Ffirst{\beta {d}}}$ \;

              \ForEach{${B\rightarrow\gamma} \in \mathcal{P}'$
              }{
                  \eIf{ ${B\rightarrow\dotInItem\gamma} \notin \Fproj{P}$
                  }{
                      add $\sItem{B\rightarrow\dotInItem\gamma}{\Delta_1}$
                      as an unmarked item to $P$ \;
                  }{
                      \If{  $(\sItem{B\rightarrow\dotInItem\gamma}{\Gamma}\in{P}$ \mbox{ and }
                            $\Delta_1\not\subseteq\Gamma)$
                      }{
                          update \sItem{B\rightarrow\dotInItem\gamma}{\Gamma}
                          to \sItem{B\rightarrow\dotInItem\gamma}{\Gamma\cup{\Delta_1}}
                          in $P$  \;

                          tag \sItem{B\rightarrow\dotInItem\gamma}{\Gamma\cup{\Delta_1}}
                          as unmarked \;
                      }
                  }
              }
          }
      }

      \Return $P$ \;
    }
\caption{\label{alg:closureOne}
         Computation of $\FclosureOne{P}$
        }
\end{algorithm}
\DecMargin{1em}

The LR(1) parsing table for $\mathcal{G}$ is constructed from an automaton,
called LR(1)-automaton,
whose states are sets of \Noneitems.
Correspondingly, function ${\Fla_i}$ is instantiated as follows.
\begin{quote}
\begin{em}
  For every final state $P$ of the LR(1)-automaton and
  for every $\sItem{A \rightarrow\beta\dotInItem}{\Delta} \in P$,
  \acapo
    ${\Fla_{LR}}\pair{P}{A\rightarrow\beta} = \Delta$.
\end{em}
\end{quote}

\noindent
LR(1)-automata are obtained by applying Alg.~\ref{alg:generic-automaton}
after:
\begin{itemize}
  \item
  using $\FclosureOne{\_}$ (see Alg.~\ref{alg:closureOne}) as
  $\Fclosure{\_}$ function, and
  \item
  taking $P_0 = \FclosureOne{\set{\sItem{S'\rightarrow\dotInItem{S}}{\set{\$}}}}$, and
  \item
  taking $\Ntmp =
          \set{\sItem{A\rightarrow\alpha{Y}\dotInItem\beta}{\Delta} \mid
               \sItem{A\rightarrow\alpha\dotInItem{Y}\beta}{\Delta}\in{P}
               \mbox{ for some } A,\alpha,\beta, \mbox{ and }\Delta}$.
\end{itemize}

\noindent
When applied to an item with projection
${A\rightarrow\alpha\dotInItem{B}\beta}$,
$\FclosureOne{\_}$ refines $\FclosureZero{\_}$
by propagating the symbols following $B$ to the closure items whose driver is $B$.
By definition, $\FclosureOne{P}$ is
the smallest set of items, with smallest lookahead-sets,
that satisfies the following equation:
\acapo
%pqtrick
  $\mbox{ }\qquad
   \FclosureOne{P} = P \cup \{  \sItem{B\rightarrow\dotInItem\gamma}{\Gamma} \mbox{ such that }
%pqtrick
    $ \acapo \phantom{$\mbox{ }\qquad\FclosureOne{P} = P \cup \{$}$
    \sItem{A \rightarrow\alpha\dotInItem{B}\beta}{\Delta}\in\FclosureOne{P} \mbox{ and }
    {B\rightarrow\gamma}\in{\mathcal{P}'} \mbox{ and }
    \Ffirst{\beta\Delta}\subseteq\Gamma
    \}.
  $

The computation of
$\FclosureOne{\set{\sItem{S'\rightarrow\dotInItem{S}}{\set{\$}}}}$
for $\mathcal{G}_1$ is detailed in the following, where we assume that
items are processed in the same order in which they are tagged as unmarked
in the collection under construction.

\begin{enumerate}
\item
  First round of {\bf{while}}
  \begin{itemize}
    \item
      $\sItem{S'\rightarrow\dotInItem{S}}{\set{\$}}$ taken as $I$
      in Alg.~\ref{alg:closureOne}, marked
    \item
      $\Delta_1 = \set{\$}$
    \item
      $\sItem{S\rightarrow\dotInItem L=R}{\set{\$}}$ added to $P$, unmarked
    \item
      $\sItem{S\rightarrow\dotInItem R}{\set{\$}}$ added to $P$, unmarked.
  \end{itemize}
\item
  Next round of {\bf{while}}
  \begin{itemize}
    \item
      $\sItem{S\rightarrow\dotInItem L=R}{\set{\$}}$ taken as $I$, marked
    \item
      $\Delta_1 = \set{=}$
    \item
      $\sItem{L\rightarrow\dotInItem *R}{\set{=}}$ added to $P$, unmarked
    \item
      $\sItem{L\rightarrow\dotInItem \Token}{\set{=}}$ added to $P$, unmarked.
  \end{itemize}
\item
  Next round of {\bf{while}}
  \begin{itemize}
    \item
      $\sItem{S\rightarrow\dotInItem R}{\set{\$}}$ taken as $I$, marked
    \item
      $\Delta_1 = \set{\$}$
    \item
      $\sItem{R\rightarrow\dotInItem L}{\set{\$}}$ added to $P$, unmarked.
  \end{itemize}
\item
  Next round of {\bf{while}}
  \begin{itemize}
    \item
      $\sItem{L\rightarrow\dotInItem *R}{\set{=}}$ taken as $I$, marked.
  \end{itemize}
\item
  Next round of {\bf{while}}
  \begin{itemize}
    \item
      $\sItem{L\rightarrow\dotInItem \Token}{\set{=}}$ taken as $I$, marked.
  \end{itemize}
\item
  Next round of {\bf{while}}
  \begin{itemize}
    \item
      $\sItem{R\rightarrow\dotInItem L}{\set{\$}}$ taken as $I$, marked
    \item
      $\Delta_1 = \set{\$}$
    \item
      $\sItem{L\rightarrow\dotInItem *R}{\set{=}}$
      updated to
      $\sItem{L\rightarrow\dotInItem *R}{\set{=,\$}}$, unmarked
    \item
      $\sItem{L\rightarrow\dotInItem \Token}{\set{=}}$
      updated to
      $\sItem{L\rightarrow\dotInItem \Token}{\set{=,\$}}$, unmarked.
  \end{itemize}
\item
  Next round of {\bf{while}}
  \begin{itemize}
    \item
      $\sItem{L\rightarrow\dotInItem *R}{\set{=,\$}}$ taken as $I$, marked.
  \end{itemize}
\item
  Last round of {\bf{while}}
  \begin{itemize}
    \item
      $\sItem{L\rightarrow\dotInItem \Token}{\set{=,\$}}$ taken as $I$, marked.
  \end{itemize}
\end{enumerate}

\begin{figure}[!h]
  \centering
  \includegraphics[width=14cm]{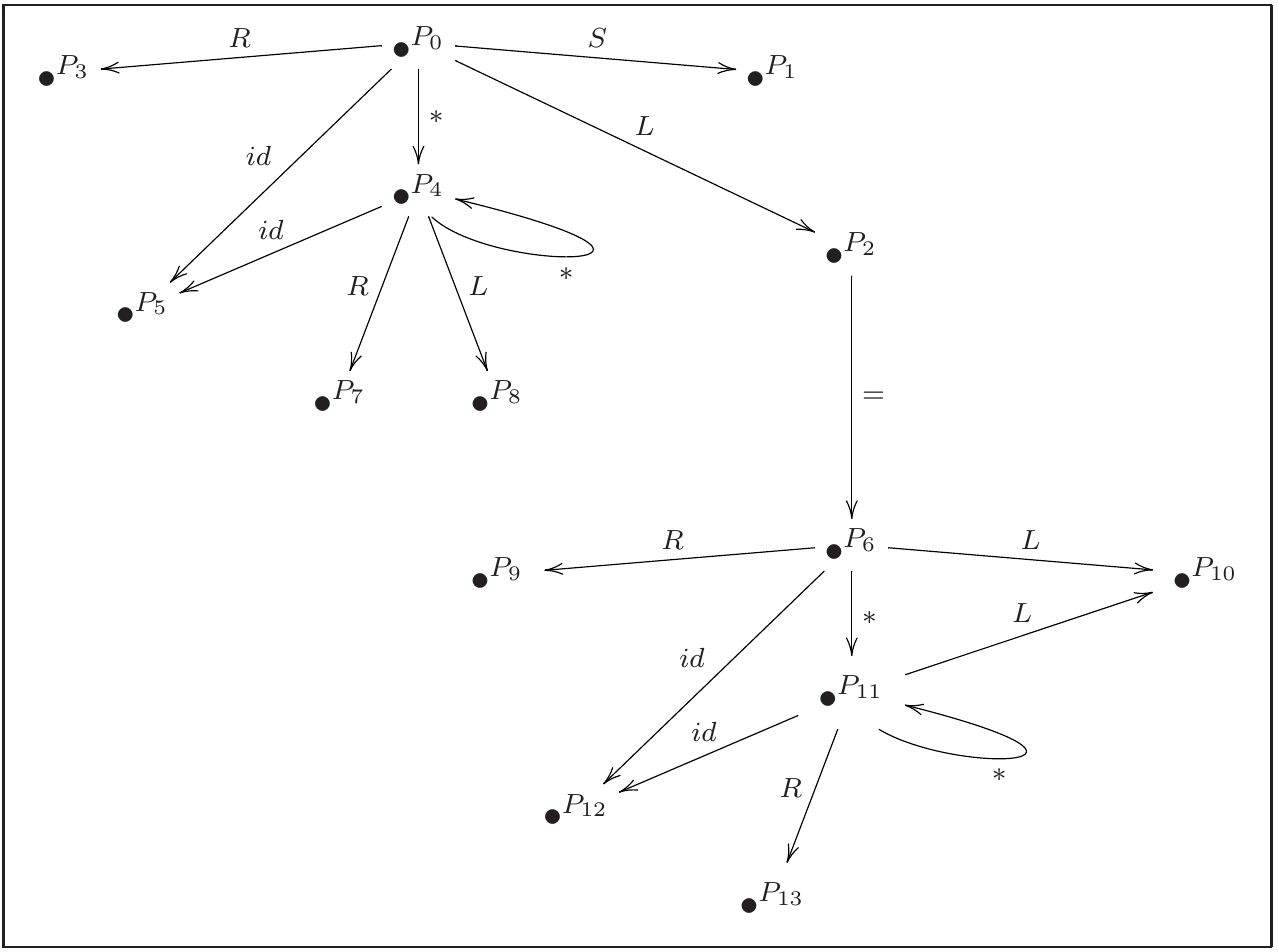}
  \caption{\label{fig:appendix_lr_pointer_automaton}
         Layout of the LR(1)-automaton for $\mathcal{G}_1$
         }
\end{figure}

\noindent
The layout of the LR(1)-automaton for $\mathcal{G}_1$ is reported in
Fig.~\ref{fig:appendix_lr_pointer_automaton}.
The accepting item is in state $P_1$.
The final states of the automaton, and the reducing items they contain,
are listed below.

\begin{center}
$\begin{array}{l@{\hspace{4ex}}l}
  \mbox{{\bf State}}      &
  \mbox{{\bf Reducing item}}
\\[.7ex]
  P_2:       &
  \sItem{R\rightarrow L\dotInItem}{\set{\$}}
\\[.7ex]
  P_3:       &
  \sItem{S\rightarrow R\dotInItem}{\set{\$}}
\\[.7ex]
  P_5 :       &
  \sItem{L\rightarrow \Token \dotInItem}{\set{=,\$}}
\\[.7ex]
  P_7 :       &
  \sItem{L\rightarrow * R \dotInItem}{\set{=,\$}}
\\[.7ex]
  P_8 :       &
  \sItem{R\rightarrow L\dotInItem}{\set{=,\$}}
\\[.7ex]
  P_9 :       &
  \sItem{S\rightarrow L=R\dotInItem}{\set{\$}}
\\[.7ex]
  P_{10} :       &
  \sItem{R\rightarrow L\dotInItem}{\set{\$}}
\\[.7ex]
  P_{12} :       &
  \sItem{L\rightarrow \Token \dotInItem}{\set{\$}}
\\[.7ex]
  P_{13} :       &
  \sItem{L\rightarrow * R \dotInItem}{\set{\$}}
\end{array}$
\end{center}

\pqcomment{
  The upper bound to the number of states of the LR(0)-automaton for $\mathcal{G}$ is
  $O(2^{n(t+1)})$, where $n$ is the size of $\mathcal{G}'$, and $t$ is the size of $T$.
  To the best of our kwnoledge, this bound is not shown to be tight.
}% end\pqcomment

\section{LALR(1) grammars}
\label{sec:lalr}

LALR(1) parsing tables are based on automata whose size is the same as the size
of LR(0)-automata.
Various algorithms achieve the same goal.

\subsection*{From LR(1)-automata}

The less efficient algorithm for the construction of LALR(1) parsing tables
is based on the use of \emph{\LRm-automata} (for LR(1)-\emph{merged}-automata).

Call $\autM$ the \LRm-automaton for $\mathcal{G}'$.
The construction of $\autM$ requires, as pre-processing,
the computation of the LR(1)-automaton for $\mathcal{G}'$, say $\autL$.
Given $\autL$, the states and the transitions of $\autM$ are defined as follows.

\begin{description}
  \item[States:]
        The states of $\autL$ are partitioned into classes of states having
        the same projection.
        Each state of $\autM$ represents one of such classes, and is defined as
        the union of the \Noneitems\ in the states of $\autL$
        belonging to the corresponding class.
  \item[Transitions:]
        If the state $M$ of $\autM$ is such that $\Fproj{M}=\Fproj{L}$,
        where $L$ is a state of $\autL$,
        and if $L$ has a $Y$-transition to $L'$,
        then $M$ has a $Y$-transition to the state $M'$
        such that $\Fproj{M'}=\Fproj{L'}$.
        We observe here that the transitions of the states of $\autL$ only depend
        on their projections.
        Hence, if a state $L$ of $\autL$ has a $Y$-transition to $L'$,
        then all the states in the same class as $L$ have $Y$-transitions
        to states in the same class as $L'$.
\end{description}

The LALR(1) parsing table for $\mathcal{G}$ is constructed from the \LRm-automaton, and
instantiating function ${\Fla_i}$ as follows.
\begin{quote}
\begin{em}
  For every final state $P$ of the \LRm-automaton and
  for every $\indexedset{\sItem{A \rightarrow\beta\dotInItem}{\Delta_j}}{j}\subseteq{P}$,
  %\acapo
    ${\Fla_{LRm}}\pair{P}{A\rightarrow\beta} = \bigcup_j \Delta_j$.
\end{em}
\end{quote}

\subsection*{From smaller automata}

The algorithm described in Sec.~4.7.5 of the international edition of~\cite{AhoLSU06}
is the so-called \yacc\ algorithm~\cite{YaccManual74}.
It uses LR(0)-automata as underlying characteristic automata for the
contruction of LALR(1) parsing tables.
The computation of the lookahead function is then based on a
post-processing phase carried on that automaton.
The post-processing phase of the \yacc\ algorithm consists in performing
closure$_1$-operations that allow the identification of \emph{generated}
lookaheads.
In various passes, the generated lookaheads are then propagated,
along the edges of the LR(0)-automaton, to the appropriate reducing items.

\bison, a well-known parser generator~\cite{BisonManual}, applies an
algorithm designed by DeRemer and Pennello~\cite{DeRemerP82}.
Like the \yacc\ algorithm, the algorithm by DeRemer and Pennello
is organized as a post-processing of LR(0)-automata.
In a nutshell, starting from the state $P$ where the reducing item
${A\rightarrow\beta\dotInItem}$ is located,
the algorithm by DeRemer and Pennello traverses the automaton
to infer which precise subset of the productions of the grammar should be
considered when computing the follow-set of $A$ for the item
${A\rightarrow\beta\dotInItem}$ in $P$.

%\input{Inputs/alg_automaton}

%%% alg_automaton
%% pag.32 del manuale: \eIf(then comment){condition}{then block}(else comment){else block}

%\IncMargin{1em}
\begin{algorithm*}%[H]
  %\LinesNumbered
  %
  %\Input{Enriched grammar $\mathcal{G}' = (V',T,S',\mathcal{P}')$}
  %\BlankLine
    $\varX_0 \assign \FnewVar$\;
    $\Cvars \assign \set{\varX_0}$\;
    $P_0 \assign \FclosureOne{\set{\sItem{S'\rightarrow\dotInItem S}{\set{\varX_0}}}}$\;
    initialize $\Cequations$ to contain the equation $\lookEquation{\varX_0}{\set{\$}}$\;
    initialize $\statesX$ to contain $P_0$\;
    tag $P_0$ as unmarked\;

    \While{ there is an unmarked state $P$ in $\statesX$ }
    {
      mark $P$ \;

      \ForEach{ grammar symbol $Y$
      }{

        \tcc{\small{Compute the kernel-set of the $Y$-target of $P$.
                   }}

        $\Ntmp \assign \emptyset$\;

        \ForEach{ $\sItem{A\rightarrow\alpha\dotInItem{Y}\beta}{\Delta}$ in $P$
        }{
          add $\sItem{A\rightarrow\alpha{Y}\dotInItem\beta}{\Delta}$ to $\Ntmp$\;
        }
        \If{$\Ntmp\neq\emptyset$
        }{
            \eIf{$\Fproj{\Ntmp}=\Fproj{\Fkernel{Q}}$ for some $Q$ in $\statesX$
            }{

              \tcc{\small{$Q$ is the $Y$-target of $P$.
                          Refine $\Cequations$ to propagate lookaheads from $P$ to $Q$.
                          }}

              \ForEach{ %pair
                             $( \sItem{A\rightarrow\alpha{Y}\dotInItem\beta}{\Delta} \in \Ntmp
                                \mathrel{,}
                                \sItem{A\rightarrow\alpha{Y}\dotInItem\beta}{\set{\varX}} \in \Fkernel{Q} )$
              }{
                  update $(\lookEquation{\varX}{\Gamma})$ to
                  $(\lookEquation{\varX}{\Gamma \cup \Delta})$ in $\Cequations$\;
              }
              $\Fgoto\pair{P}{Y}\assign{Q}$\;
            }{

              \tcc{\small{Generate the $Y$-target of $P$.
                         }}

              \ForEach{ $\sItem{A\rightarrow\alpha{Y}\dotInItem\beta}{\Delta} \in \Ntmp$
              }{
                $\varX \assign \FnewVar$\;

                $\Cvars \assign \Cvars\cup\set{\varX}$\;

                enqueue $(\lookEquation{\varX}{\Delta})$ into ${\Cequations}$\;

                replace $\sItem{A\rightarrow\alpha{Y}\dotInItem\beta}{\Delta}$
                by $\sItem{A\rightarrow\alpha{Y}\dotInItem\beta}{\set{\varX}}$ in $\Ntmp$\;
              }
              $\Fgoto\pair{P}{Y}\assign\FclosureOne{\Ntmp}$\;

              add $\Fgoto\pair{P}{Y}$ as an unmarked state to $\statesX$ \;
          }
        }
      }
    }
\caption{\label{alg:automaton}
         Construction of the symbolic automaton
         %for $\mathcal{G} = (V,T,S,\mathcal{P})$
        }
\end{algorithm*}
%\DecMargin{1em}

Below, we describe an algorithm based on the construction of specialized
\emph{symbolic characteristic automata}~\cite{c_2016_MFCS}.
The states of these automata are sets of \emph{symbolic items}, which have
the same structure as \Noneitems.
The lookahead-sets of symbolic items, however,
can also contain elements from a set \NvarSet\ which is disjoint from $V'\cup\set{\$}$.
Elements of \NvarSet\ are called \emph{variables} and are ranged over by
$\varX,\varXp,\ldots$.
In what follows, we use
$\Delta,\Delta',\ldots,\Gamma,\Gamma',\ldots$
to denote subsets of $\NvarSet\cup T\cup\set{\$}$.
Also, we let $\Fground{\Delta} = \Delta\cap(T\cup\set{\$})$.
%%%, and $\Fvar{\Delta} = \Delta\cap\NvarSet$.
Moreover, we assume the existence of a function \FnewVar\ which returns a
fresh symbol of \NvarSet\ at any invocation.
The definitions of initial, accepting, kernel, closure, and reducing items
are extended to symbolic items in the natural way.
Also, functions $\Fproj{\_}$ and $\Fkernel{\_}$
are overloaded to be applied to sets of symbolic items.

Variables are used to construct on-the-fly a symbolic version of the
\LRm-automaton.
In every state $P$ of the symbolic automaton,
the lookahead-set of kernel items is a singleton set
containing a distinguished variable, like,
e.g. $\sItem{A \rightarrow\alpha{Y}\dotInItem\beta}{\set{\varX}}$.
On the side, an equation for $\varX$ collects all the contributions to the
lookahead-set of ${A \rightarrow\alpha{Y}\dotInItem\beta}$ coming from
the items with projection ${A \rightarrow\alpha\dotInItem{Y}\beta}$ which
are located in the states $Q_i$ with a $Y$-transition to $P$.
When a new state $P$ is generated and added to the current collection,
$\FclosureOne{\_}$ symbolically propagates to the closure items
the lookaheads encoded by the variables associated with the kernel items of $P$.
\pqcomment{%
  Given that $\FclosureOne{\_}$ behaves like the identity function when applied to
  a set of reducing items, there is no need to care about such propagation
  from reducing items.
  Hence, kernel reducing items are treated differently from the non-reducing items
  in the kernel.
  The lookahead-sets of kernel reducing items are thought of as accumulators, and
  directly record the contributions coming from the corresponding items in
  the states that have a transition to $P$.
}% end \pqcomment
When the construction of the symbolic automaton is over,
the associated system of equations over variables is resolved to compute,
for every variable $\varX$,
the subset of $T\cup\set{\$}$ that is the actual value of $\varX$,
denoted by $\Fval{\varX}$.
The evaluation of variables, in turn, is used to actualize lookahead-sets.
In particular, function ${\Fla_i}$ is instantiated as follows.
\begin{quote}
\begin{em}
  For every final state $P$ of the symbolic automaton and
  for every $\sItem{A \rightarrow\beta\dotInItem}{\Delta} \in P$,
  \acapo
    ${\Fla_{LALR}}\pair{P}{A\rightarrow\beta} =
      \Fground{\Delta} \cup
      \bigcup_{\varX\in\Delta}\Fval{\varX}$.
\end{em}
\end{quote}

\begin{figure}%[!t]
 \centering
$\begin{array}{ll@{\hspace{14ex}}l}
  \mbox{{\em State}}      &
  \mbox{{\em Items (kernel in purple)}}   &
  % equations
  \Cequations
\\[1ex]
  P_0: &
  \bcode{\sItem{S'\rightarrow\dotInItem S}{\set{\varX_0}}} &
                                                            \lookEquation{\varX_0}{\set{\$}}
\\
  & \sItem{S\rightarrow\dotInItem L=R}{\set{\varX_0}}
\\
  & \sItem{S\rightarrow\dotInItem R}{\set{\varX_0}}
\\
  & \sItem{L\rightarrow\dotInItem *R}{\set{=,\varX_0}}
\\
  & \sItem{L\rightarrow\dotInItem \Token}{\set{=,\varX_0}}
\\
  & \sItem{R\rightarrow\dotInItem L}{\set{\varX_0}}
\\[1ex]
  P_1:       &
  \bcode{\sItem{S'\rightarrow S\dotInItem}{\set{\varX_1}}} &
                                                            \lookEquation{\varX_1}{\set{\varX_0}}
\\[1ex]
  P_2:       &
  \bcode{\sItem{S\rightarrow L \dotInItem  =R}{\set{\varX_2}}} &
                                                            \lookEquation{\varX_2}{\set{\varX_0}}
\\
  & \bcode{\sItem{R\rightarrow L\dotInItem}{\set{\varX_3}}} &
                                                            \lookEquation{\varX_3}{\set{\varX_0}}
\\[1ex]
  P_3:       &
  \bcode{\sItem{S\rightarrow R\dotInItem}{\set{\varX_4}}} &
                                                            \lookEquation{\varX_4}{\set{\varX_0}}
\\[1ex]
  P_4:       &
  \bcode{\sItem{L\rightarrow * \dotInItem R}{\set{\varX_5}}} &
                                                            \lookEquation{\varX_5}{\set{=,\varX_0}\cup\set{\varX_5}\cup\set{\varX_7}}
\\
  & \sItem{R\rightarrow\dotInItem L}{\set{\varX_5}}
\\
  & \sItem{L\rightarrow\dotInItem *R}{\set{\varX_5}}
\\
  & \sItem{L\rightarrow\dotInItem \Token}{\set{\varX_5}}
\\[1ex]
  P_5 :       &
  \bcode{\sItem{L\rightarrow \Token \dotInItem}{\set{\varX_6}}} &
                                                            \lookEquation{\varX_{6}}{\set{=,\varX_0}\cup\set{\varX_5}\cup\set{\varX_7}}
\\[1ex]
  P_6:       &
  \bcode{\sItem{S\rightarrow L= \dotInItem R}{\set{\varX_7}}} &
                                                            \lookEquation{\varX_7}{\set{\varX_2}}
\\
  & \sItem{R\rightarrow\dotInItem L}{\set{\varX_7}}
\\
  & \sItem{L\rightarrow\dotInItem *R}{\set{\varX_7}}
\\
  & \sItem{L\rightarrow\dotInItem \Token}{\set{\varX_7}}
\\[1ex]
  P_7 :       &
  \bcode{\sItem{L\rightarrow * R \dotInItem}{\set{\varX_8}}} &
                                                            \lookEquation{\varX_8}{\set{\varX_5}}
\\[1ex]
  P_8 :       &
  \bcode{\sItem{R\rightarrow L\dotInItem}{\set{\varX_9}}} &
                                                            \lookEquation{\varX_9}{\set{\varX_5}\cup\set{\varX_7}}
\\[1ex]
  P_9 :       &
  \bcode{\sItem{S\rightarrow L=R\dotInItem}{\set{\varX_{10}}}} &
                                                            \lookEquation{\varX_{10}}{\set{\varX_7}}
\\[1ex]
\end{array}$
  \caption{\label{fig:appendix_pointer_states}
         Symbolic automaton for $\mathcal{G}_1$: content of states, and of $\Cequations$
         }
\end{figure}

%\input{Inputs/alg_reduced_system}

%%% alg_reduced_system

%%%%%%%%%%%%%%%%%% per togliere la virgola ai commenti \tcc*[h] invece che \tcc*[r]
\begin{algorithm}
  %\LinesNumbered
  %\BlankLine

    inizialize $\RCvars$ and $\RCequations$ to $\emptyset$ \;

    \While{ $\CequationsBypassing$ not empty
    }{

        $\lookEquation{\varX}{\Delta} \assign \Fdequeue{\CequationsBypassing}$ \;

        \tcc{\small{For $j>i$, if $\lookEquation{\varX_j}{\{\varX_i\}}$
                    then $\varX_j$ is in the same class as $\varX_i$.
                    The same holds of $\lookEquation{\varX_j}{\{\varX_i,\varX_j\}}$,
                    i.e. up to self-reference.
                    }}

        \eIf{ $\Delta\setminus\set{\varX} = \set{\varXp}$
        }{

            $\Cclass{\varX} \assign \Cclass{\varXp}$ \;
        }{

            $\Cclass{\varX} \assign \varX$ \;
            add ${\varX}$ to $\RCvars$ \;

        }

    }
    %%%%%
    %\BlankLine
    %%%%%
    \tcc{\small{Clean up the defining equations of the variables in $\RCvars$.
                Use only representative variables on right-sides,
                and remove self-references.
                }}

    \ForEach{ $\varX \in \RCvars \mbox{ such that }
               \lookEquation{\varX}{\Delta} \in \CequationsBypassing$
    }{

        update each $\varXp$ in $\Delta$ to $\Cclass{\varXp}$  \;

        add $\lookEquation{\varX}{\Delta\setminus\set{\varX}}$ to $\RCequations$ \;
    }
    %%%%%

\caption{\label{alg:reduced_system}
         Reduced system of equations $\RCequations$ for the variables in $\RCvars$
        }
\end{algorithm}

\begin{figure}%[!t]
 \centering
  \includegraphics[width=3cm]{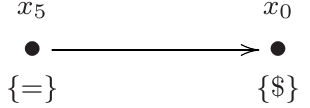}
  \caption{\label{fig:graph_reqs_pic}
         Dependency graph for the reduced system of equations obtained
         from $\Cequations$ in Fig.~\ref{fig:appendix_pointer_states}
         }
\end{figure}

%\input{Inputs/alg_traverse_DeRemer}

%%% alg_traverse_DeRemer

\begin{algorithm}%[H]
  \BlankLine
    \ForEach{$\varX$ represented by a vertex of $DG$
    }{

      \tcc{\small{Variable to be used as index for strongly connected components (SCC).}}

      $scc(\varX) \assign 0$ \;
    }
    \ForEach{$\varX$ in represented by a vertex of $DG$
    }{
      \If{$scc(\varX) = 0$
      }{
          $\Ftraverse{\varX}$ \;
      }
    }
    \BlankLine

    where

    \BlankLine

    \Fun{$\Ftraverse{\varX}$
    }{
      push $\varX$ onto stack $S$ \;
      $depth \assign$ number of elements in $S$ \;
      $scc(\varX) \assign depth$ \;
      $\Fval{\varX} \assign init(\varX)$ \;
      \ForEach{$\varX'$ such that there is an edge in $DG$ from ${\varX}$ to ${\varX'}$
      }{
          \If{$scc(\varX') = 0$
          }{
             $\Ftraverse{\varX'}$
          }

            \tcc{\small{When the entry vertex of an SCC is found backwards,
                        all the vertices in the SCC take the same index
                        as the index of the entry vertex.
                        }}

            $scc(\varX) \assign min(scc(\varX),scc(\varX'))$ \;

            $\Fval{\varX} \assign \Fval{\varX} \cup \Fval{\varX'}$ \;
      }
      \tcc{\small{The entry vertex of an SCC finds all the vertices of the SCC
                  on top of the stack (the entry vertex is the deepest),
                  it assigns its value to all the vertices in the SCC and cleans
                  the stack up. (Note: Any leaf of a tree is an SCC.)
                  }}

      \If{ $scc(\varX) = depth$
      }{
          \Repeat{ $pop(S)=\varX$
          }{
              $scc(top(S)) \assign \infty$ \;
              $\Fval{top(S)} \assign \Fval{\varX}$ \;
          }
      }
    }
\caption{\label{alg:traverse_DeRemer}
         Computation of the values for the variables in the dependency graph $DG$
        }
\end{algorithm}

\noindent
The procedure for collecting all the elements needed to set up
the LALR(1) parsing table consists in the following steps.

\begin{enumerate}
  \item
    Construct the symbolic automaton by applying Alg.~\ref{alg:automaton},
    and get the set $\Cvars$ of variables generated for the construction,
    and the list $\Cequations$ of equations installed for those variables.

    The application of Alg.~\ref{alg:automaton} to $\mathcal{G}_1$ results in a
    symbolic automaton with the same layout as that of its LR(0)-automaton
    (Fig.~\ref{fig:appendix_pointer_automaton}).
    The content of the states of the symbolic automaton, and the associated
    system of equations $\Cequations$
    are both reported in Fig.~\ref{fig:appendix_pointer_states}.

  \item
    Set up a graph $DG$ for the computation of the actual values of variables.

    $DG$ is the dependency graph of the reachability relation embedded by
    the definining equations.
    Each vertex of $DG$ represents one of the variables occurring as left-side
    of an equation.
    If the equation to be represented for $\varX$ is
    $\lookEquation{\varX}{\Delta}$,
    then the vertex for variable $\varX$ has an outgoing edge to each of
    the vertices for the variables in $\Delta$.
    Also, the vertex for $\varX$ is associated with the inital value
    $init(\varX)=\Fground{\Delta}$.
    
    Computational efficiency can be gained by operating on a dependency graph
    smaller than that induced by $\Cequations$.
    This can be achieved after the following observations.

    Let
    $\CvarsReducing\subseteq\Cvars$
    be the set of  variables associated with reducing items, and let
    $\CvarsBypassing = \Cvars \setminus \CvarsReducing$.
    By construction, all the variables in $\CvarsReducing$ cannot propagate
    any further, and rather act as accumulators.
    In fact, each of the variables in $\CvarsReducing$ occurs in $\Cequations$
    only once, as left-side of its defining equation.
    Then, to solve the system of equations, it is sufficient to compute the
    values of the variables in $\CvarsBypassing$.
    Once these values are known, for each $\varX_i\in\CvarsReducing$
    such that $\lookEquation{\varX_i}{\Delta_i}$ is in $\Cequations$,
    we can set
    \begin{equation}
    \label{eq:CvarsReducing}
      \Fval{\varX_i} =
      \Fground{\Delta_i} \cup \bigcup_{\varX\in\Delta_i}\Fval{\varX}.
    \end{equation}

    The second observation is that the variables in $\CvarsBypassing$ can
    be partitioned into equivalence classes, so that it is enough
    to evaluate one variable per class.
    Let $\CequationsBypassing$ be obtained from $\Cequations$ by removing
    the equations for the variables in $\CvarsReducing$.
    To get the partition of the variables in $\CvarsBypassing$,
    we run Alg.~\ref{alg:reduced_system} over $\CequationsBypassing$.
    Alg.~\ref{alg:reduced_system} returns a reduced system of equations
    $\RCequations$ which define the variables in $\RCvars\subseteq\CvarsBypassing$.
    Also,
    every variable $\varX\in\CvarsBypassing$ is associated with
    a class representative, denoted by $\Cclass{\varX}$.

    As for the running example, the set $\CvarsBypassing$ for
    the symbolic automaton of $\mathcal{G}_1$ is given by
    $\set{\varX_0,\varX_1,\varX_2,\varX_5,\varX_7}$.
    The application of Alg.~\ref{alg:reduced_system} to the corresponding
    set of equations $\CequationsBypassing$ results in the
    reduced system $\RCequations$ shown below, and the induced
    dependency graph is drawn in Fig.~\ref{fig:graph_reqs_pic}.

    \begin{center}
    $\begin{array}{l@{\hspace{7ex}}l@{\hspace{7ex}}l@{\hspace{7ex}}l}
        \CequationsBypassing &
        \Cclass{\varX} &
        \RCvars &
        \RCequations
      \\
        \lookEquation{\varX_0}{\set{\$}} &
        \varX_0 &
        \varX_0 &
        \lookEquation{\varX_0}{\set{\$}}
      \\
        \lookEquation{\varX_1}{\set{\varX_0}} &
        \varX_0 &
      \\
        \lookEquation{\varX_2}{\set{\varX_0}} &
        \varX_0 &
      \\
        \lookEquation{\varX_5}{\set{=, \varX_0, \varX_5, \varX_7}} &
        \varX_5 &
        \varX_5 &
        \lookEquation{\varX_5}{\set{=, \varX_0}}
      \\
        \lookEquation{\varX_7}{\set{\varX_2}} &
        \varX_0 &
      \\
    \end{array}$
    \end{center}

  \item
    Compute the values of the variables in $\RCvars$.

    This is obtained by running Alg.~\ref{alg:traverse_DeRemer} on $DG$.
    Alg.~\ref{alg:traverse_DeRemer}, by DeRemer and Pennello~\cite{DeRemerP82},
    is an adaptation of a depth-first visit for finding strongly connected
    components~\cite{Tarjan72}.
    In particular, Alg.~\ref{alg:traverse_DeRemer} specializes an
    algorithm presented in~\cite{EveK77} for the efficient
    computation of the reflexive and transitive closure of
    arbitrary relations.
    Briefly, the values associated with the farthest nodes
    are accumulated with the values of the nodes found along the way back to
    the origin of the path.
    The visit of the graph is organized in such a way that strongly connected
    components are recognized on-the-fly, and, as due, each vertex in the
    connected component is associated with the same value.

    Running Alg.~\ref{alg:traverse_DeRemer} on the graph in
    Fig.~\ref{fig:graph_reqs_pic}, we get
    $\Fval{\varX_0}= \set{\$}$, and
    $\Fval{\varX_5}=\set{=,\$}$.

  \item
    Compute the values of the variables in $\Cvars\setminus\RCvars$.
    
    First, actualize the values of the variables in
    $\varX\in\CvarsBypassing\setminus\RCvars$ using
    equation~(\ref{eq:CvarsBypassing}) below.
    \begin{equation}
    \label{eq:CvarsBypassing}
      \Fval{\varX}=\Fval{\Cclass{\varX}}.
    \end{equation}
    Then compute the values of the variables in $\varX\in\CvarsReducing$
    by using equation~(\ref{eq:CvarsReducing}).
    
    In the case of the symbolic automaton of $\mathcal{G}_1$, given the
    values computed for ${\varX_0}$ and for ${\varX_5}$, we get
    $\Fval{\varX_1}=
     \Fval{\varX_2}=
     \Fval{\varX_3}=
     \Fval{\varX_4}=
     \Fval{\varX_7}=
     \Fval{\varX_{10}}=
     \set{\$}$, and
    $\Fval{\varX_6}=
     \Fval{\varX_8}=
     \Fval{\varX_9}=
    \set{=,\$}$.
    
\end{enumerate}

%%%  BIBER %%%%%%\printbibliography
%\bibliography{parsingBibNoDoiShort}

\end{document}